# Blockchain and human episodic memory

Cho SH[1]*, Cushing CA[2]*, Patel K[3], Kothari A[3], Lan R[3], Michel M[4], Cherkaoui M[2], Lau H[1,2,3,5,6]

1 Department of Psychology, University of Hong Kong, Pokfulam Road, Hong Kong
2 Department of Psychology, UCLA, Los Angeles, 90095, USA
3 Harmony.One, California, 94087, USA
4 Department of Philosophy, Sorbonne-Université, Paris, France
5 Brain Research Institute, UCLA, Los Angeles, 90095, USA
6 State Key Laboratory of Brain and Cognitive Sciences, University of Hong Kong, Pokfulam Road, Hong Kong

*These authors contributed equally to this work.

Correspondence: SeongHahCho@gmail.com

**<u>Abstract</u>**

We relate concepts used in decentralized ledger technology to studies of episodic memory in the mammalian brain. Specifically, we introduce the standard concepts of linked list, hash functions, and sharding, from computer science. We argue that these concepts may be more relevant to studies of the neural mechanisms of memory than has been previously appreciated. In turn, we highlight that certain phenomena studied in the brain, namely metacognition, reality monitoring, and how perceptual conscious experiences come about, may inspire development in blockchain technology too, specifically regarding probabilistic consensus protocols.

## Introduction

Decentralized ledger technology, also known as blockchain, promises to transform how information is stored and shared on the internet. Despite the financial magnitude of the excitement surrounding the technology (Young 2018), the fact that blockchain technology could also bring new conceptual developments in other scientific fields such as neuroscience remains underappreciated. In this article, we surmise that certain concepts related to blockchain can be applied to studies of the human brain, specifically about how we store consciously experienced information from an autobiographical viewpoint, a capacity known as episodic memory. In turn, we suggest that borrowing concepts from neuroscience could also be useful for the development of blockchain technology.

Our view is not that the brain implements a blockchain *per se*. But nor do we think that the brain exactly implements ideal Bayesian (i.e. optimal probabilistic) inference – and yet, this hasn't stopped Bayesian models from providing useful insights for understanding brain functions, and from generating meaningful formal hypotheses to stimulate experimentation (Peters and Lau 2015; K. J. Friston, Harrison, and Penny 2003; Lake, Salakhutdinov, and Tenenbaum 2015; Ma et al. 2006; Beck et al. 2008; Körding and Wolpert 2004). It is also informative to know how the brain deviates from ideal Bayesian norms (Morales et al. 2015; Maddox and Bohil 1998; Rahnev and Denison 2018). We hope to learn similar lessons by seeing how blockchain may provide useful analogies as well as disanalogies for understanding episodic memory mechanisms in the brain. In general, borrowing concepts from the latest information technologies to form theoretical frameworks has proven fruitful in neuroscience (Marr and Poggio 1979; Van Surdam Graham 1989; Miller, Galanter, and Pribram 2013; Minsky 1974; Gazzaniga 2004; Baddeley and Weiskrantz 1995).

Unfortunately, perhaps in part because of the financial speculation surrounding blockchain technology (and one of its applications, Bitcoin), some may consider much of what has been written on the topic to be unduly exaggerated. Although skepticism towards analogies between fashionable technologies and brain processes is warranted, we believe that careful and qualified analogies might still prove useful. We will address this issue throughout, and specifically discuss the limits of the analogy between blockchain and brain processes in the last section.

## What is blockchain?
In simple terms, blockchain is a method for storing information in a decentralized system, to guarantee that the concerned parties agree on how and what new information is added, and that none of them can unilaterally tamper with previous records. As an intuitive analogy, we can think of blockchain as keeping a permanent ledger up 'in the sky', so to speak, for everyone to see (when certain conditions are met). Roughly, this ledger is open, immutable, and 'append-only'; that is, one can only *add* new information to the blockchain, but not erase, or distort, previously registered information.

More specifically, in a typical simple blockchain design, there are numerous 'nodes' (i.e. computers) holding duplicated copies of a ledger (Figure 1). The ledger is like a book in which each new page is called a 'block'. To add a new block, some cryptographic puzzle depending on the content of previous blocks must be solved.The incentive structure is set up such that when node runners solve the cryptographic puzzle before others do, they win the competition to add a new block, and thus a token; that's why node runners are also called 'miners' (as if they were mining valuable coins).

Because the cryptographic solution for writing a new block depends on the content of previous blocks, if one tampers with such content, one's solution will no longer be accepted by others. This process of **consensus,** i.e. that one's proposed cryptographic solution needs to be approved by others through some voting mechanism (Bano et al. 2017), prevents malicious miners from participating in the agreed competition, and ensures that the content of the ledger is immutable.

One of the most well-known uses of blockchain is probably Bitcoin (Nakamoto 2008). In essence, Bitcoin uses the method described above to record financial transactions in the currency of the token for mining. But the same decentralized ledger technology can also be used for many other purposes (Underwood 2016; Kuo, Kim, and Ohno-Machado 2017; Ølnes, Ubacht, and Janssen 2017), wherever a record needs to be kept open and trustworthy.

In this paper, we surmise that some properties of human episodic memory could be analogical to some of the properties of blockchain. Notably, the autobiographical memory of our conscious experiences is a somewhat linear chain of subjective episodes stored in our brains, which are themselves systems supported by the distributed activities of billions of neurons – in a 'decentralized' way, just as a blockchain is. Although our memory is not perfect, the capacity to voluntarily 'delete' (i.e., forget) an episode is limited (Anderson and Hanslmayr 2014). Moreover, although episodic memory encoding is achieved through noisy and spatially distributed neuronal mechanisms, our episodic memory remains relatively stable over time. For instance, we don't confuse our own childhood with that of a character we have read in a novel. Despite occasional errors – which one would expect in any biological systems – it is not trivial at all to completely erase a specific memory by tampering with neurons in a specific brain region (Abdou et al. 2018; Ryan et al. 2015; Roy et al. 2017).

We suggest that closer examination of how far the analogies between blockchain and episodic memory can go could be a fruitful exercise. To do so, we present key concepts used in decentralized ledger technology, and see to what extent they may inform neuroscience research, specifically in the area of episodic memory.

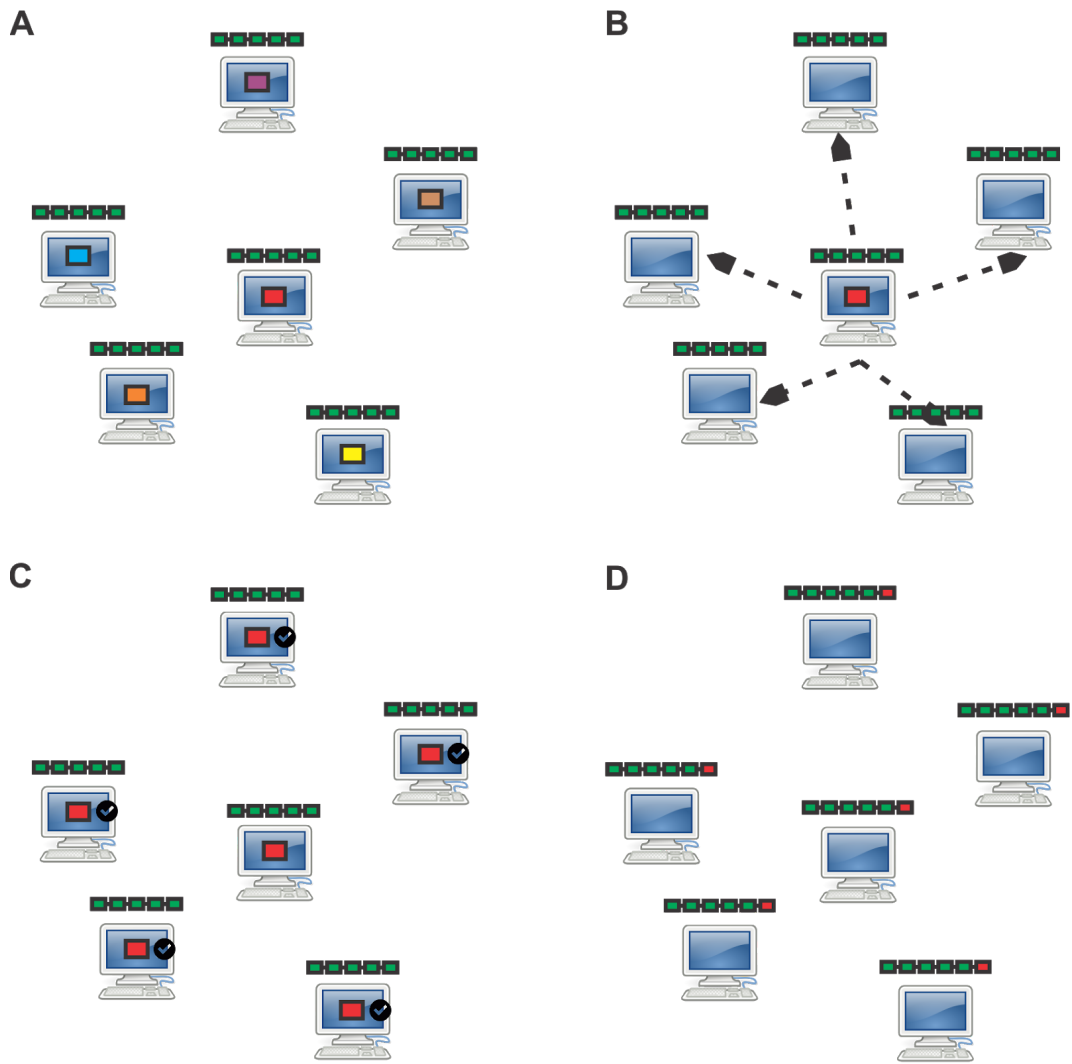

**Figure 1. Process used by Blockchains to append new blocks. A**) Each node in the blockchain network work independently to solve a cryptographic puzzle. The puzzle in Bitcoin involves producing a hashed output containing specific properties. The first node to solve the puzzle has the opportunity to add a block containing a list of transactions to all the nodes in the network. In doing so, the node receives a reward in return. Note that each node has a copy of the ledger, hence the term "distributed ledger". **B**) The node that first completes the puzzle will broadcast both the solution and the block contents to all other nodes in the network. The remaining nodes will stop solving the cryptographic puzzle upon receiving a solution. **C**) Blocks that receive the solution and block contents of the winning node begin the process of transaction verification. Each node in the network will verify the solution and the transaction list to ensure that the solution is correct and all transactions are possible and valid. **D**) Once this has been verified, consensus has been reached and the nodes will add the block to their currently held ledger. The process repeats again from A for the next block.

**Blockchain concepts for neuroscience**

*(1) Data structures: Linked list vs. array*

In computer science, it is well known that one can store and access information more or less efficiently by using different kinds of **data structures**. A simple and common data structure is an **array**, where values are stored in a specific order, such that each value is accessible by referring to its specific position in the array. However, arrays have one critical shortcoming: their size must be *pre-allocated* in memory. That is, regardless of whether arrays are full or empty, they occupy the same amount of space in the computer's memory.

To avoid this pitfall, one can appeal to another common type of data structure: the **linked list**. Each item in a linked list stores two things: a symbol addressing a stored value and the address of the next item in the linked list. This simplifies insertion and extension operations, but a major drawback is that one cannot directly index a linked list by element position. To access the fifth item, the system must access the first item in the linked list and then move forward through links in the list until it reaches the fifth item, rather than having direct access to the fifth item.

Human episodic memory seems better modeled by a linked list rather than array, at least if one is to choose only between these two options. Indeed, the dynamic flexibility afforded by a linked list makes it a sensible choice for a data structure representing the uncertain and wide-ranging structure of human autobiographical memory. Despite the temporal structure of episodic memory, we cannot really access it in the serialized fashion representative of an array. Indeed, compared to other storage cognitive systems, such as visual working memory, information stored in episodic memory does not seem to be serially accessible: people generally have a much easier time recalling the third item from a row of recently viewed items than they do recalling the third episode of their week. Cued with a random date like October 21, 1997, we're not so good at remembering the episodes that occurred on that date, unless it happens to be within proximal context to a landmark event that has coincidentally seared that date into memory. To this extent, introspection on episodic memory seems akin to a series of linked episodes, with one episode leading to the next and old episodes being called up by contextual cues bringing us back in time. This is why retracing one's footsteps and recalling the details of an episode surrounding a misplaced item can serendipitously bring back the memory of where the item has been placed (Godden and Baddeley 1975; Ranganath 2010; Horner and Burgess 2014; Horner et al. 2015; Chan et al., 2017).

Perhaps the biggest limitation of the array data structure as a model of episodic memory is its finite predefined size. The data structure occupies a constant amount of space in the computer's memory, regardless of whether it is full or empty: it cannot grow and shrink on command like a linked list. Once again, arrays are a better model for something like visual working memory, the capacity of which is relatively constant across testings within an individual (Luck and Vogel 2013). We can easily find a limit to how many items one can keep track of in visual working memory, but finding the limit of episodic memory is no small task. It seems unlikely that the brain could pre-allocate a large enough array for episodic memory during development, thereby wasting valuable neural capability with an inflexible data structure.

If information is stored in episodic memory by using a data structure similar to a linked list, one central question is to know how stored values are defined in those linked lists, namely, how sequences of our conscious lives are segmented into autobiographical events with definite boundaries. Buzsáki & Tingley (2018) have proposed that the hippocampus is a sequence generator – carving and concatenating events and experiences into sequences, which is akin to a linked list. Studies of event boundaries in episodic memory suggest that different brain regions have preferred timescales to segment information into meaningful events (Baldassano et al., 2017; Hasson et al., 2015), and that the timescales increase as information flows away from primary sensory regions (Murray et al., 2014). Events are then aggregated within longer timescales regions before eventually prompting the hippocampus to store an event in episodic memory (Baldassano et al. 2017). In this framework, the hippocampus could partition and store the informational content between two event boundaries as an **episodic block**, thereby segmenting episodic memory into meaningful events to facilitate storage and retrieval of relevant information.

The hypothesis according to which episodic memory could be segmented into episodic blocks seems to be supported by behavioral studies (Radvansky and Zacks 2017). For example, Ezzyat & Davachi (2011) used narrative stimuli to alter the event structure within a story by including sentences like 'a while later', indicating an event change. They observed that the ability of participants to recall information across events decreased compared to information within event even if the story content was unchanged.

In addition, physiological studies of the hippocampus also seem to support the hypothesis that episodic memory is structured in episodic blocks. Studies using fMRI have shown that hippocampus activity increases with the occurrence of a perceived event boundary (Baldassano et al. 2017; Ben-Yakov and Henson 2018; Ben-Yakov, Rubinson, and Dudai 2014). This increase in hippocampus activity could correspond to the reinstantiation of a prior event to foster long term storage of the event that is being committed to memory (Sols et al. 2017). There is also evidence that memory encoding and access occurs holistically across related items (Horner et al. 2015). That is, contents with associative relationships are functionally bound together, as would be expected by a blocked structure of episodic memory. Therefore, evidence suggests that the hippocampus segments continuous experience into discrete episodic blocks.

To complete the blockchain analogy, these discrete episodic blocks must be linked. One hypothesis posits that links between memories in the brain are stored inherently in the neural memory structure. Through synaptic plasticity, memories could be carved into ensembles of neurons called engrams (Tonegawa et al. 2015). During memory formation, the likelihood of a given neuron to be recruited into a new engram is dependent upon its excitability: neurons in an excitable state are more likely to be recruited into an engram, and neurons recruited into an engram, in turn, become more excitable (Rashid et al. 2016; Park et al. 2016).

Building on this observation, some researchers have tried to explain the link between certain episodes in time as a result of the ongoing adult neurogenesis in the dentate gyrus (Aimone, Wiles, and Gage 2006). Dentate gyrus neurons have different response properties at different stages of their life cycle, with younger, more excitable neurons being more likely to be recruited into a pattern for memory during engram formation. This suggests that memories formed close in time recruit overlapping populations of neurons into their respective engrams, providing a structural link between temporally co-occurring episodes. Therefore, episodes from a similar point in time (on the scale of weeks as neurons mature), share a greater pattern overlap compared to episodes more distal in time. As such, when an old memory is cued, memories from the same period become more retrievable (Aimone, Wiles, and Gage 2006). With similar CREB-modulated mechanisms controlling engram formation on the hour time scale (Lisman et al. 2018), it seems plausible that similar mechanisms carry linkage over longer time scales. The rare and ongoing neurogenesis in dentate gyrus provides both the physical substrate and properties necessary to keep an autobiographical linked list growing over a lifetime.

*(2) Hash functions*

Blockchains are more than a linked list. They also make critical use of **hash functions** to keep information secure and trustworthy. In our view, borrowing the concept of hash functions from decentralized ledger technology could also be useful for understanding human memory functions.

The basic concept of hash functions is common in computer science. Essentially, it is a method for taking an input message of any size and turning it into a string of a fixed length. Some of the most commonly used hash functions have a property called collision resistance. With a collision resistant hash function, it is extraordinarily unlikely for two messages to produce the same output string. Message hashes are created by feeding the message into an algorithm that changes some internal variables based on the input. These internal variables are highly sensitive to each part of the message. The process is iterated over the entire message to produce an output that is highly sensitive to small changes in the message.

There are various real-life applications of hash functions. For example, hashing can be employed to check if email messages have been corrupted while being delivered. This is done by sending the hashed string alongside the email. By sending both, the recipient can verify the message integrity by checking the hash string: mismatched message/hash pairs would indicate that the message was modified. Changes in the original message generates changes in the hash string due to a property of hash functions. Likewise, applying a hash function to a search query term produces a hash that can be used as an index to efficiently access data on a database in a data structure known as a hash table.

Blockchains use hash functions for two purposes: first, as a cryptographic puzzle the aim of which is to find a particular hash string and second, by hashing the content of a block, comparisons of the resulting hash can verify consistency of content between distributed copies of blocks. The ability of hash functions to generate fixed-size references to data in a collision resistant manner is particularly relevant to episodic memory. We further explore this relationship below.

One could consider pattern separation in the hippocampus as a process analogous to hashing for episodic memory. Pattern separation consists in taking memory representations, say the appearance of an orange and a tangerine, and orthogonalizing how the two memories are stored. This prevents interference between two similar memories that might cause incorrect recall of a memory to a cue, such that, for instance, an orange is not mistaken for a tangerine and vice versa. In the following paragraphs, the relationship between pattern separation and hashing will be explored by emphasizing shared properties between the two processes.

One property of hashing is the reduction of a message into a smaller string. Similarly, pattern separation sparsifies, or reduces population activity, as memory representations are tracked from the cortex to the hippocampus. Through the process of sparsification, semantically similar representations shift from being encoded by large, overlapping percentages of entorhinal cortex neurons to smaller, independant percentages in the dentate gyrus (Knierim and Neunuebel 2016). Sparsification also occurs at CA3, as sparse dentate gyrus activity combined with low dentate gyrus outputs to CA3 results in small populations of CA3 cells per memory (McNaughton and Morris 1987; O'Reilly and McClelland 1994; Treves and Rolls 1992). Thus, from the entorhinal cortex to the CA3, memory representations are mapped onto a much smaller population of cells, akin to the transformation of message to hash.

Sparsity is also central for another property shared between blockchain and episodic memory: collision resistance. Collision resistance in episodic memory requires minimal overlap between representations in addition to unique representations. This contrasts with hash functions, for which unique representations are sufficient. Hash functions achieve collision resistance by algorithms that act pseudo-randomly on the input. In episodic memory, pattern separation achieves it through sparsification and random recruitment of memory encoding neurons (Rolls 2008). Sparse activity is important to separate representations as it reduces the risk that a single neuron holds representations of two overlapping memories. In addition, the sparse dentate gyrus activity and few connections to the CA3 has a randomizing effect on which CA3 cells are recruited to represent a specific memory (Rolls 2016). This further orthogonalizes the CA3 engram cells between memories, ensuring collision resistance.

Collision resistance has also been empirically evaluated. Studies using gradually changing environments found that the dentate gyrus and CA3 cells recruit independent populations of neurons to orthogonalize the spatial representation of similar contexts (S. Leutgeb et al. 2004; J. K. Leutgeb et al. 2007). In addition, studies looking at the input-output relationship of the dentate gyrus found sharp decorrelation of dentate gyrus activity with shifting cues compared to its input regions, the medial entorhinal cortex and the lateral entorhinal cortex (Neunuebel et al. 2013; Neunuebel and Knierim 2014). Pattern separation is also seen in human fMRI studies. For example, in a study by Bakker et al. (2008), subjects were presented with an initial image of an object followed by variable images of the same object. Comparison of activity between the repeated presentation of the initial image and variable image revealed large differences in the dentate gyrus/CA3 activity. Combined, these findings suggest a collision resistant encoding that separates similar memory representations.

Hashing memory episodes has several functional advantages. Hashing prevents memory representations with similar cued contexts from entangling, thereby preventing the recall of a false memory instead of the correct memory. Our daily experiences often contain unchanging spatial and environmental cues, separated only by time. For instance, if you consistently park a car in the same parking structure, some mechanism must exist to differentiate the current location of the car from its previous locations. Thus, our memory abilities not only require a large storage capacity, but also a means of storing, indexing, and reading a memory that incorporates collision resistance across incoming and past memories. Pattern separation serves this purpose in the hippocampus.

One speculative advantage of using hash functions in conjunction with a linked list is that it incorporates a means of checking for corruption or damage. Let's assume that a memory block contains the memory representation plus the hash of another block in a memory sequence. This, in turn, is repeated in prior blocks all the way to the first block within the memory linked list. Pattern separation must then occur over the combination of memory representation *and* hash when adding new blocks. Therefore, the output of dentate gyrus/CA3 could be a unique hash, dependent not only on the memory content but also on the history of all prior hashes as well. Similar to message tampering detection, the memory hash can, in principle, be used to verify that all recalled contents belong to the memory chain. If alteration of the memory linked list occurs, it would completely transform the hash signaling at fault and possibly deny recall of any subsequent episodes. In doing so, this mechanism could help protect against possible attacks on memory integrity, including the insertion of false memories or recall with missing content (Loftus and Pickrell 1995).

*(3) Sharding*

One issue currently challenging the development of blockchain technology is that of scalability (Zibin Zheng et al. 2016; Vukolić 2016), i.e. how to maintain speed and efficiency while allowing many users and transactions to happen. Because in traditional models all nodes process every transaction, this often results in extremely slow transaction times. The two largest implementations, Bitcoin and Ethereum, can process up to ~7 and ~20 transactions per second, respectively, compared to Visa at ~2000 transactions per second. In addition, delay to confirm a transaction takes up to 1 hour (Bitcoin) or 15 minutes (Ethereum; "Blockchain Confirmations - What Are They And Why Do They Matter?" 2019). This severely limits the practical usefulness of blockchains as they stand.

Scalability issues mostly stem from the following problems: the cost of submitting and broadcasting to the blockchain network and the cost of reaching consensus in traditional blockchain designs. Cost is defined as the necessary computational and communication resources. First, it stems from the limited number of transactions contained in a block plus the non-trivial times to mine and add a block. Second, when a node commits a transaction to the ledger, that transaction must then be validated by all the remaining nodes in the network before a consensus is reached; this is how trustworthiness of the ledger is established. This delay scales logarithmically as each node confers a small latency as they are added to the network.

One solution to the above problems is **sharding** (Cattell 2011; Chen and Guestrin 2016). Currently this is an active area of research, and a strategy implemented by several major groups involved in blockchain technology (Kokoris-Kogias et al. 2017; Durov 2017; Luu et al. 2016).

Sharding has been widely applied in database systems well before blockchain was invented (Ceri, Negri, and Pelagatti 1982; Agrawal, Narasayya, and Yang 2004). It is a method devised to address scaling limitations as an alternative to using a more powerful CPU, more RAM, more disk space, and so on. A sharded database will distribute data across multiple servers, and it will distribute incoming requests across those servers. When serving a request, each server is responsible for ensuring that its response is consistent with the responses of all other servers. From there, each shard is duplicated in multiple copies. Because each shard is maintained at a certain size, the relevant copies also do not become unmanageably large.

Consider this example to understand how sharding can help. Large online retailers, like Amazon, could store user data like shipping addresses, credit card info, etc., on a single server. If this one server dealt with all incoming traffic, due to hardware limitations, the rate of incoming requests could easily surpass the databases' ability to process requests. By splitting the database across multiple instances according to some principle (i.e. across geographic locations), the database becomes much more efficient as workload is shared.

In blockchains, sharding takes on another layer of complexity as copies of the transactions need to be distributed. This results in challenges to remain decentralized while maintaining quick transaction times and a high level of security. Currently, there is no agreed upon method of sharding, as this is still a region of ongoing research (Kokoris-Kogias et al. 2017; Durov 2017; Luu et al. 2016). Below, we consider sharding as a relevant analogy for how episodic memory is stored within the brain.

One particular instance showing some evidence of sharding in the brain is the circuit involved in contextual fear conditioning. Contextual fear conditioning is a paradigm used to study the circuit underlying fear memories which typically involves pairing a context with a mild electric shock to induce a fear association (Curzon, Rustay, and Browman 2011). Fear responses can then be assessed using behavior (e.g. freezing response in mice).

Neuroscience studies using this paradigm reveal several key regions that demonstrate sharding of episodic memory. A well explored property of episodic memory is that the storage of a memory exists across multiple brain regions. Memories are decomposed into its constituent parts (e.g. car and traffic accident) and the engrams are located in regions specific to the engram content. One study traced the connections from the hippocampus and entorhinal cortex during contextual fear conditioning experiments (Kitamura et al. 2017). It revealed projections to the prefrontal cortex and the basolateral amygdala among other regions, both of which hold engrams necessary for the fear response (Tonegawa et al. 2015; Zelikowsky et al. 2014; Rashid et al. 2016; Ohkawa et al. 2015). Similar to geographic-based sharding, the storage site of engrams in the prefrontal cortex and basolateral amygdala take advantage of the functional roles these regions play. In the contextual fear conditioning circuit, the prefrontal cortex is known to play a role in integrating temporal, contextual, and predictive cues and is thought to be where some of these relationships might be stored (Gilmartin, Balderston, and Helmstetter 2014; Euston, Gruber, and McNaughton 2012; Hyman 1982; Touzani, Puthanveettil, and Kandel 2007). On the other hand, the basolateral amygdala is a well-studied region associated with emotional processing (Janak and Tye 2015) and has been implicated in encoding the qualitative goodness (valence) of some object (Belova et al. 2007; Paton et al. 2006; Schoenbaum, Chiba, and Gallagher 1999). Therefore, the basolateral amygdala engram might take advantage of such processing. Combined, the prefrontal cortex and basolateral amygdala engrams offer a complete fear memory while being located in functionally relevant regions.

There's also evidence of dissociated storage across memories. Several studies have used human fMRI to explore brain activity as subjects recalled memories originating over a range of time (Smith and Squire 2009; Woodard et al. 2007; Douville et al. 2005). They found that activation across regions occurred in an age-dependent manner, such that the recall of older memories resulted in different activation patterns when compared to younger memories. Such dissociation may reflect changes in the storage as the memory becomes increasingly schematized or integrated into a pre-existing framework of knowledge.

In addition to partitioned storage, blockchains are protected against attacks or offline nodes by storing copies of each shard distributed across multiple nodes. Similarly, in episodic memory, within the regions comprising the contextual fear conditioning circuit, there appears to be distributed copies of the memory. More importantly, these sharded copies can elicit a compensatory response that triggers the recovery of the fear response. One experiment using CA1 region inhibition showed compensatory activity in the anterior cingulate cortex that recovers remote contextual fear memories (Goshen et al. 2011). In addition, direct optogenetic activation of the retrosplenial cortex produced a fear response during hippocampal inactivation (Cowansage et al. 2014). Observation of downstream amygdala and entorhinal cortex activity found that the naturally cued and optogenetic induced responses were indistinguishable. This suggests a hippocampus independent circuit for memory recall that is indistinguishable from a hippocampus dependent circuit. Furthermore, there also exists evidence for distributed copies of content within a sharded engram. Lesions of the basolateral amygdala found that the bed nucleus of the stria terminalis could compensate in producing near-normal fear responses (Poulos et al. 2010). This compensatory mechanism was absent in a combined basolateral amygdala/bed nucleus of the stria terminalis lesion.

**Neuroscience concepts for blockchain**

*(1) Consensus without global broadcast?*

To the extent that the analogies between blockchain and neuroscience are useful, we believe that both domains can inspire each other. In particular, if episodic memory in the mammalian brain is *somewhat* like a blockchain, one natural question arises as to what determines how a new memory block is created. In terms of brain functions, this concerns the mechanisms for conscious perception (Dehaene, Lau, and Kouider 2017; Lau and Rosenthal 2011). In terms of blockchain, this corresponds to the consensus protocol (Z. Zheng et al. 2017). Both are currently active areas of research.

Not all information available in the brain enters episodic memory. The brain is a distributed system, i.e., a network of neuronal processes, in which some representations remain unconscious and unreportable (Marcel 1983; Dehaene, Lau, and Kouider 2017). Our remembered stream of consciousness, on the other hand, is relatively coherent, as a serial chain of experiences, despite the underlying parallel processes. Early on, researchers thought that episodic memory and conscious experiences are intimately linked (Tulving 1985). Intuitively, the process that selects what enters consciousness can be considered as the gating mechanism for what gets written into episodic memory. That is, we don't tend to consciously recollect events that we have not consciously experienced in the first place; false but vivid memories should be rare.

How does the brain select what enters consciousness, amongst competing representations? One dominant idea in the literature is that there is a global broadcasting mechanism, called the "global workspace", implemented by a fronto-parietal network with long range and dense connections with other sensory and peripheral regions (Dehaene & Changeux, 2011; Dehaene, Lau, and Kouider, 2017). When a representation successfully enters this global workspace, the corresponding information is broadcasted throughout the entire network, resulting in conscious access to the information, and its (possible) encoding into episodic memory.

Taking this as a analogy for blockchain, this form of consensus may correspond to early designs where all nodes in the network are involved. However, as consciousness research is an emerging field, there are alternative proposals.

One such proposal is based on the observation that the brain has a metacognitive system that serves the purpose of internal monitoring (Lau and Rosenthal 2011). In particular, this system has probably evolved because the same sensory neurons can be excited for different reasons. For instance, some neurons fire (i.e. send out impulses as computational signals) when we consciously see faces. The same neural representations are also activated when we imagine faces (O'Craven and Kanwisher 2000), try to remember a certain face (Khader et al. 2005), or dream about faces (Horikawa et al. 2013). In fact, these neurons also routinely fire spontaneously, in a way that is similar to 'noise'. And yet, at least in the last scenario there is no conscious experience of seeing a face whatsoever. We also don't tend to confuse imagination or dreams with normal seeing. As such, neuronal circuits within the prefrontal and parietal cortices could support a certain internal reality monitoring system allowing us to automatically infer the causes of our sensory representations (Simons et al. 2008). Note that even though this monitoring mechanism seems somewhat centralized, it could nonetheless be distributed in different cortical areas, to some extent, in terms of implementation. Based on lesion studies we also know there should be at least some degree of redundancy, in the sense that it is not easy to completely abolish this function with a single localized damage (Fleming et al. 2014).

Incidentally, this hypothesis on the mechanisms for selecting representations for conscious processing also borrows from an analogy from machine learning (Goodfellow et al. 2014). Neural networks for pattern recognition (e.g. so-called Deep Learning models) can benefit from *predictive coding*. That is, instead of just taking information in passively, they make top-down 'hypotheses' about the world, to be tested against observations. This way, they are more robust. Trouble is, training these generative models often require too much time and data. One solution to this is to use *generative adversarial networks* (GANs), which have been hailed as the "coolest idea in deep learning in the last 20 years" (Castelvecchi 2017).

In GANs, one network is trained to generate, say, pictures of cats mimicking real cats. Another network, known as the Discriminator, is trained to distinguish between these generated cats images and real cats images. The two networks are pitted against each other within a competitive point system, such that the generating network is trained by trying to "fool" the Discriminator which, in turn, gets better at spotting the generating network's mistakes. This creates a virtuous circle of self-correction, making learning very efficient for both networks. In this context, the internal monitor needed for deciding what enters consciousness in the brain could be similar in architecture to the Discriminator: it is developed so as to tell if an early sensory representation is caused by (and therefore truthfully represents) an event or object in the world right now.

Of course, neither Discriminators in GANs nor reality monitoring systems in the brain are perfect: they make mistakes. These mistakes are not so desirable in blockchains, where complete trustworthiness of the data is the ultimate goal. However, as researchers start to push for the time efficiency of consensus protocols, they too realize that it is difficult to achieve perfect consensus with a large decentralized system, even if sharding methods are applied (Kokoris-Kogias et al. 2017). As such, one proposed solution to achieve efficiency is the so-called "trust-but-verify" method, in which transactions of relatively small amounts are approved quickly without invoking a full consensus procedure. If they turn out to be illegitimate after all, one can revisit them and follow up accordingly.

In a sense, perhaps the human mind works the same way: consciousness is just a quick but not bulletproof mechanism of selecting what representations to trust. We tend to believe in what we consciously see, but if upon checking, things are not as they seem, we do ultimately revise our thoughts and beliefs. After all, mistakes do happen, as one can hallucinate (Silbersweig et al. 1995) or just make regular perceptual errors.

To the extent that the analogy holds, perhaps one lesson for the development of consensus protocol is that we can adopt a strategy similar to GANs within the 'trust-but-verify' framework. Although training neural networks is a slow exercise, demanding large amount of both data and computational power, once a network is trained its application is relatively fast and straightforward. One may think that training a network to detect disingenuine, i.e., malicious nodes, is difficult (Athalye, Carlini, and Wagner 2018; Carlini and Wagner 2016, 2017; Athalye and Carlini 2018), as this may require a large amount of data (and faces other challenges). But, taking an adversarial framework, one could set up a competing network to simulate the behavior of these malicious nodes. This would be similar to fire drills, which can perhaps stimulate the successful training of a Discriminator network to quickly decide if the behavior of a certain node is trustworthy. This is of course not to say that it would be an easy problem to solve (Athalye, Carlini, and Wagner 2018; Carlini and Wagner 2016, 2017; Athalye and Carlini 2018), but if the brain as a decentralized system does it this way too, perhaps there is a moral for development of blockchain technology as well.

*(2) Distributed representations for scalability*

One glaring weakness of traditional blockchains, mentioned earlier, is their inability to scale with increased transaction throughput. Block and transaction verification prior to consensus is a time-consuming process that, if improved, could vastly expand the practical usability of distributed ledger technologies.

Interestingly, neurons in the lateral amygdala (a region where engrams are found) receive projections that activate ~70% of the total neuronal population present during memory encoding (Repa et al. 2001; Johansen et al. 2010). However, observations during recall of a memory reveal activations of a much smaller sub-population of the initial responsive cells (Rumpel et al. 2005; Reijmers et al. 2007). This is also observed across the entorhinal cortex to the dentate gyrus (Rolls 2016). Thus, there seems to be a shared premise between blockchains and memory encoding, where the transactions (or memories) are overexpressed across the networks (or neurons) and can be pared down while remaining immutable and fault tolerant. Fortunately, it seems that our brains have already developed some organizational principles that might inspire blockchain technologies to achieve the same reduced but adequate representations. This section attempts to apply these principles to distributed ledger technologies to help address issues of scalability.

Our memories are encoded and recalled using distributed representations (Wixted et al. 2014). Namely, memories are encoded by small populations of cells instead of either a whole population or single cells. For instance, some cells in a car engram ensemble activated during recall might also be activated during recall of a boat or train. These representations have overlapping but overall distinct populations.

It is plausible that a distributed representation-like organization could be applied to blockchains to reduce the transaction verification time when validating a block. Currently, all transactions are verified by all nodes or, in a sharded blockchain, all transactions are verified by a subset of nodes. However, even sharding could result in latencies if the block sizes are large. One possibility would be to further distribute the transactions across random nodes, such that a block with 10,000 transactions might split the transactions into 1,000 transaction segments and redundantly distribute verification across a subset of the available nodes. This could also occur within a shard. Parallelization results in reduced verification time when compared to the current implementations. Thus, a distributed representation like redistribution of transactions could help increase the scalability of blockchains.

However, a possible caveat of a distributed representation implementation is that it introduces what is called the double spending problem, namely, it creates a loophole amid multiple representations, allowing a malicious user to get away with creating two conflicting transactions (Nakamoto 2008). Distribution of transactions results in the separation of a user's transactions across multiple nodes. If there's no overlap in nodes, the two conflicting transactions may be successfully verified even if the user lacks resources for both transactions. Our previous sections, *Data structures* and *Hash functions,* provide some insight into resolving this issue. To prevent double spending, one possible solution is to implement a semi-structured node selection algorithm based on hippocampal episodic encodings for how states (i.e., segments of the complete ledger) are distributed across nodes. Concepts from the recruitment of neurons using both structured and random selection to an engram are borrowed. The remaining paragraphs will discuss the selection process that recruits neurons to an engram applied to blockchains.

Earlier in the *Data structures* section, we discussed CREB based linkage of memories in the CA1 based on temporal proximity. Persistently elevated CREB levels following prior memory encoding recruits the same neurons for a following memory, linking the engrams together (Lisman et al. 2018). This phenomenon, observed in CA1, could inspire the organization of transaction verification within the nodes of a network. In a distributed representation implementation, the transaction verification can occur such that transactions within a defined temporal window originating from a user are assigned to an overlapping network of nodes. This would guarantee that all double spent transactions undergo verification by a node that has seen both transactions and is capable of flagging the double spend and rejecting it. One possible suggestion for a temporal window can be determined by the necessary time to confirm a transaction which, in the case of Bitcoin, can take up to an hour. This method could prevent any attempts to double spend prior to confirmation.

In the *Hash functions* section, we mentioned that the sparse but robust projections from the dentate gyrus to CA3 also contribute to orthogonalized representations (Rolls 2016; Knierim and Neunuebel 2016). Some proposed sharded implementations use random assignment of nodes for verification to prevent coordination across rogue nodes. However, random recruitment has undesirable consequences as well. Indeed, this could slow down verification if a node is randomly assigned a disproportionately high number of verification requests, in which case, verifying blocks slows until this overloaded node can complete its queue. Again, inspiration from the hippocampus can be applied to node selection to create orthogonalized nodes for verification. Neurogenesis results in orthogonalization where adult-born dentate gyrus cells are highly excitable and thus more likely to be recruited for the encoding of a memory (Aimone, Wiles, and Gage 2006). The same could be applied to recruitment of nodes in addition to random selection. Newer nodes can be given priority and therefore better distribute verification across nodes to avoid bottlenecking. This has the added benefit of allowing new nodes to become better integrated into the network early on.

Another possibility is to find a way of tracking the last time a node was used for verification. Nodes that have been left unused for verification are prioritized to be recruited for upcoming block verifications, thereby orthogonalizing node selection.

In sum, using a hippocampal inspired, structured as well as random, recruitment of nodes could offer some advantages to current implementations of distributed ledger technology. Using distributed but structured node selection helps contain similar transactions together while decreasing latency during consensus. Random selection also adds some redundancy in storage to aid in keeping the transaction history secure in addition to adding robustness during consensus. Together, this creates a balance between efficiency, dispersion of transaction information, and robustness in consensus.

## Hypotheses for episodic memory and blockchain

This section suggests several hypotheses to further explore the similarities between blockchains and our episodic memory systems. As blockchain development and modeling of autobiographical memories share similar obstacles, we predict a convergence of the subsequent solutions. Therefore, evaluating these hypotheses should be fruitful toward understanding episodic memory.

Our first hypothesis is that disrupting the linkage between two memories should result in the additional disruption of in-between memories (Figure 2A). A unique feature of linked lists is that they store the temporal sequence across contents as a consequence of block order. Blocks placed next to each other contain stored memories that occurred consecutively in time. As such, the temporal relationship between two events exists in the order of storage. The temporal relationship between some event A and event D can be determined by the order of encoding, contained in the series of linked events between events A and D. Such time-related information across blocks resulting from block sequence should be lost if the links were to be  severed. In this hypothetical case, position between events relative to each other would no longer be available. Thus, one can predict that in a linked-list-like organization, if the memories of event A and event D are disrupted, memories located between events A and D (e.g. events B and C) are likely to be disrupted as well. In turn, this should result in impaired recall of the order of events B and C relative to other events (e.g. events E and F).

On the other hand, in an array, temporal information should be stored explicitly. In the case of an array like organization, such disruptions to events A and D are likely to be isolated to the these memories. In the above example, recall of events B and C would be unaffected.

It needs to be qualified that changes leading to disruption and loss in memories is unlikely to resemble strictly a linked-list-like or an array-like organization. However, there are advantages to a linked-list-like implementation, and our hypothesis tests where the similarities and dissimilarities lie across both organizations. Our experiences and prior studies suggest that long-term memories are more robust to attack (e.g. electroconvulsive therapy; Squire et al. 1975). Therefore, in long-term memories, disrupting events A and D may not affect events B and C. However, the temporal relationships across memories likely occurs alongside other, sequentially organized relationships. By affecting the linkage, other associations between events (e.g. causal) could be distorted. For instance, if event A is the losing of a phone and event B is the buying of a phone, a causal relationship exists between event A and event B. That is, A and B are linked not only through a relation of temporal succession, but also by a causal relation. This could influence recall used to probe event order, confounding the source of disruption. Indeed, incorrect recall, in this case, could indicate either that memory of the *causal* relation has been impaired by the manipulation, or memory of the temporal sequence, or both. In any case, testing this hypothesis could provide insight into the stored structure of an episodic memory sequence.

Our second hypothesis is that recovery of overwritten memories is possible, even when cueing with a stimulus does not elicit recall of the memory (Figure 2B). In the implementation for blockchains, linked lists are updated through adding or appending content to the end of the list. They are not updated by the changing or overwriting of previously existing content. Therefore, prior memories should still persist in the brain, even if the memories cannot be recalled naturally. Classical fear conditioning considered that memories still persist during extinction. However, studies found that through the process of reconsolidation, memories can be disrupted beyond recovery. Related research have used optogenetic activation of engram cells to recall forgotten memories (Abdou et al. 2018; Ryan et al. 2015; Roy et al. 2017). However, these experiments induce forgetting by preventing protein synthesis associated with consolidation. These experiments mostly induce the encoding of novel contexts, instead of a previously experienced context. Therefore, recovery is inconclusive in determining memory structure, as both a linked list or array could result in optogenetic recovery. Instead, the ideal design relies on replacing the content of a pre-existing memory.

One possible means of evaluating this is to induce false memories about a prior event. Prior research has demonstrated that false memories can be induced upon repeated exposure to misinformation (Loftus and Pickrell 1995). Here, inducing a false memory should result in the cued recalling of the false memory and not the original memory. In other words, the person should only have conscious recollection of the false event. However, if through testing of the original memory, some original memory content can still be recovered, this would suggest that the original memory is still preserved in the brain, even if it cannot be naturally recalled. In an array-like organization, the original memory would have been replaced with the false memory, thus resulting in poor performance when testing for the original memory. On the other hand, if the memory is amended downstream, as expected in a linked-list-like implementation, performance on the recall of the true memory should still be higher than chance. This could help us decide whether outdated information still persists or not.

However, it should be qualified that Blockchains lack the equivalent of 'false memories'. Consensus protocols prevent the addition of fraudulent transactions in current Blockchain implementations, thus failing to provide insight on how false memories are added to the memory chain. However, one possible Blockchain phenomena that allows for different blocks at the same position concurrently is through forking. In Bitcoin, miners compete to add the next block to the chain, sometimes resulting in the simultaneous addition of multiple blocks. In the terminology of Bitcoin, the chain "forks", resulting in two alternative chains competing for legitimacy. Similarly for memories, in an sequence of events, A to B to C, the false memory induction could result in a forking such that both event B and the false event B' follow from event A and lead back to event C (Figure 2C). This allows for both memories to persist and has the added benefit of preserving the order of events. In addition to evaluating permanence of memories, further experiments could be designed to evaluate how false memories are added to the memory chain.

Our third hypothesis is that attacks to an event's contents result in a failure to recall, to protect the integrity of the memory (Figure 2C). Here, an attack is defined as any change to the memory that occurs in the absence of stimuli. For instance, if a person had a stroke, the associated loss of memory engram cells would be considered an attack. However, changes in memory due to experienced events would not be considered an attack. One useful feature of hash functions for episodic memory is the ability to reflect and detect even minor changes to its contents.

One possible method for evaluating this hypothesis is by changing an engram to replicate an attack on the memory trace. If the recall of the attacked event is reduced and affects neighboring memories, this would provide evidence for our hypothesis. Previous studies from Horner et al. (2015) have shown that all associated event content is recalled simultaneously regardless of task relevance. That is, multiple event elements encoded together are all recalled when just one element is recalled. Therefore, we can evaluate the blockchain-like defensive process by recreating an attack through tagging and lesioning the cells associated with the memory trace. If such mechanisms occur, lesioning the cells storing some part of event A, should result in the impaired recall of the remaining content of event A. However, if such mechanisms are absent, these effects should be isolated only to the memory held by the cell and not to the neighboring elements.

Nonetheless, while in blockchain protection against attacks occurs through hash functions, our hypotheses is not that, in the brain, protection against attacks occurs only through the same circuitry as pattern separation. Instead, our hypothesis is an attempt to understand if the protective mechanisms exist in a manner similar to what would be expected in blockchains. If such experiments are designed carefully, they could yield insights into how the brain monitors memory integrity. However, to our knowledge, this has yet to be explored.

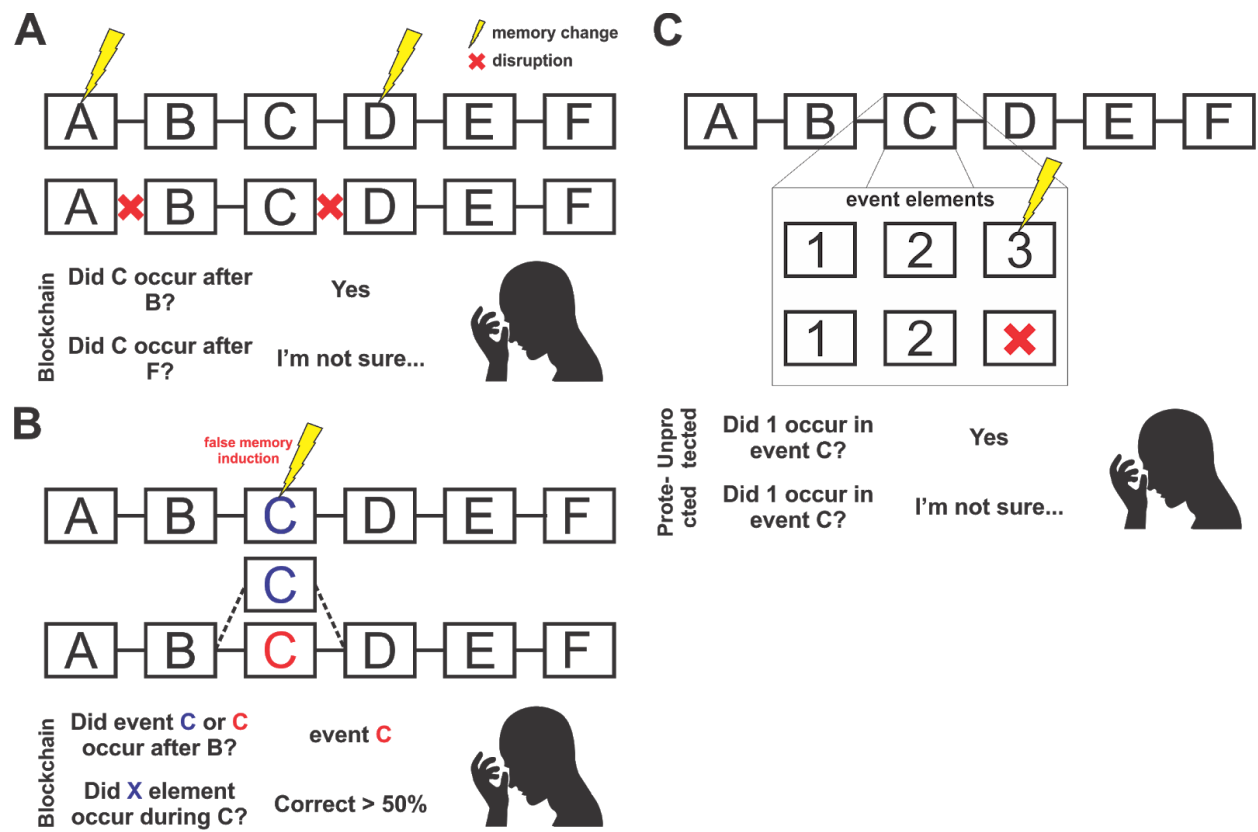

**Figure 2. Hypotheses to explore the relationship between Blockchains and episodic memory. A**) Blockchains store the temporal information in the block order. Therefore if our episodic memory is similar, event B is understood to occur after event A due to the sequence of encoding. To test this hypothesis, the memories of two events are disrupted. If this occurs, then the linkage of these events to other events are also disrupted. Consequently, this should cause the event order to be disrupted between events B and C such that it's relative position to other events could not be recalled. Therefore, indicating that the order of events is stored in the relative position on a continuous chain such that disruption of the chain results in consequences outside of the disrupted memory. **B**) Blockchains can only append blocks to an existing linked list. Therefore, new content can only be added. If episodic memory is append only, this would suggest that outdated memories still persist even though they cannot be recalled. One possible

method is through inducing a false memory altering the memory of a prior event. This should result in the subject recalling the induced false memory of the prior event. Then the subject is tested on some element of C of the original memory. If the subject performs above chance, even without conscious recollection of the prior memory, this would be evidence for persistence. Therefore, suggesting the persistence of episodic memories. **C**) Our third hypothesis predicts that altering a single element of an event results in the failure to recall the entire event. Here, event C, composed of elements 1 to 3, has undergone attack to alter the contents of element 3. If there exists a process that protects content integrity as seen in Blockchains, the event C block should be flagged as being corrupt. This manifests in the subject being unable to recall all elements of event 3. However, if our episodic memory lacks these integrity preserving mechanisms, only element 3 would result in altered recall.

**Caveats & Concluding Remarks**

Blockchain has received considerable interest and is seen as a solution to many problems faced in various industries. It has found promising application in health care, open science, but perhaps most notably in finance (Underwood 2016; Kuo, Kim, and Ohno-Machado 2017; Ølnes, Ubacht, and Janssen 2017; Tapscott and Tapscott 2017). By June 2018, the blockchain market was estimated at $256 billion (CoinMarketCap) with considerable potential for growth (Young 2018).

Here we explored the utility of blockchain as an analogy for understanding the brain mechanisms underlying episodic memory. However, analogies only work so far as they can take you. We believe that such analogies are useful to the extent that we recognize the disanalogies as well – all models are wrong, but some felicitous falsehoods are "true enough" to be useful (Box 1976; Elgin 2017). Our stance is that we should not throw the baby out with the bathwater.

In general, fields like psychology, artificial intelligence, neurophysiology, have all benefited from such analogies, especially in the case of computational neuroscience and vision research. Had David Marr and vision scientists like Norma Graham not borrowed and applied concepts from electrical engineering, our relevant textbook knowledge today would be vastly different (Marr and Poggio 1979; Van Surdam Graham 1989). We emphasize that the very premise of cognitive neuroscience involves taking concepts from information technology (Sutton, Barto, and Williams 1992; Daw et al. 2006; Watkins and Dayan 1992; Rumelhart, Hinton, and Williams 1986). This is the historical origin of the field. Recently, advances in artificial intelligence and deep learning have also very much both borrowed from as well as inspired neuroscience (Hassabis et al. 2017). For instances, they have allowed for the investigation of the emergence of grid-like representations (Banino et al. 2018) and dopaminergic activity and its influence on prefrontal cortex (Wang et al. 2018).

At times, these analogies may seem somewhat far fetched, and skepticism may be warranted (Bowers and Davis 2012). By applying Bayesian models, researchers have been able to delineate how neural activity respond to situations of uncertainty in an approximately optimal manner, leading to the Bayesian coding hypothesis (Knill and Pouget 2004). But the empirical support for the hypothesis is currently under some debate (Rahnev and Denison 2018). A somewhat related proposal has led to a so-called unified brain theory (Friston 2010). It posits that the brain instantiates a hierarchical generative model of the world, and minimizes its entropy which is characterized as free energy based on formulations akin to thermodynamics. This so-called free energy principle has proven popular. To this date, this paper has been cited over 2500 times (Friston 2010). In broader contexts, the analogy application of broadstroke concepts from physics and information theoretic analysis to biology also has a long history (Schrödinger, 1944).

With caution, we have likewise argued that concepts in contemporary blockchain technology can be applied to advance our understanding of mechanisms supporting episodic memory. As in similar cases, we ultimately believe that the usefulness of the analogy can only be evaluated by the extent to which it stimulates experiments and provides meaningful interpretations for empirical data. However, we note that blockchain technology is very much in an early stage of development. As such, new ideas have greater potential to inspire cross semination between the disciplines of neuroscience and decentralized ledger technology still.

**<u>Acknowledgement</u>**

We thank Tommaso Furlanello, Bilwaj Goankar, and Nicco Reggente for helpful discussion.